\documentclass[aps, prx,twocolumn,reprint,superscriptaddress]{revtex4-1}

\usepackage{graphicx}
\usepackage{amsmath}
\usepackage{setspace}
\usepackage{braket}
\usepackage{natbib,mathrsfs,accents}
\usepackage[dvipsnames]{xcolor}
\definecolor{myblue}{named}{MidnightBlue}
\definecolor{mygreen}{RGB}{0,120,0}
\usepackage[colorlinks=true,citecolor=myblue,linkcolor=myblue,urlcolor=myblue]{hyperref}
\usepackage{siunitx}

\usepackage[caption=false]{subfig}
\begin{document}

%\begin{frontmatter}

\title{Space-borne quantum memories for global quantum communication}

\author{Mustafa G\"{u}ndo\u{g}an*}
\affiliation{Instit\"{u}t f\"{u}r Physik, Humboldt-Universit\"{a}t zu Berlin, Newtonstr. 15, Berlin 12489, Germany}
\email{mustafa.guendogan@physik.hu-berlin.de}
\author{Jasminder S Sidhu} 
\affiliation{SUPA Department of Physics, University of Strathclyde, John Anderson Building, 107 Rottenrow East, Glasgow, G4 0NG, UK}
\author{Victoria Henderson} 
\affiliation{Instit\"{u}t f\"{u}r Physik, Humboldt-Universit\"{a}t zu Berlin, Newtonstr. 15, Berlin 12489, Germany}
\author{Luca Mazzarella}
\affiliation{SUPA Department of Physics, University of Strathclyde, John Anderson Building, 107 Rottenrow East, Glasgow, G4 0NG, UK}
\author{Janik Wolters} 
\affiliation{Deutsches Zentrum f\"{u}r Luft- und Raumfahrt e.V. (DLR), Institute of Optical Sensor Systems, Rutherfordstr. 2, 12489 Berlin, Germany}
\affiliation{Technische Universit\"{a}t Berlin, Institut f\"{u}r Optik und Atomare Physik, Str. des 17. Juni 135,
10623 Berlin, Germany}
\author{Daniel KL Oi} 
\affiliation{SUPA Department of Physics, University of Strathclyde, John Anderson Building, 107 Rottenrow East, Glasgow, G4 0NG, UK}
\author{Markus Krutzik}
\affiliation{Instit\"{u}t f\"{u}r Physik, Humboldt-Universit\"{a}t zu Berlin, Newtonstr. 15, Berlin 12489, Germany}

\begin{abstract}
Global scale quantum communication links will form the backbone of the quantum internet. However, exponential loss in optical fibres precludes any realistic application beyond few hundred kilometres. Quantum repeaters and space-based systems offer to overcome this limitation. Here, we analyse the use of quantum memory (QM)-equipped satellites for quantum communication focussing on global range repeaters and Measurement-Device-Independent (MDI) QKD. We demonstrate that satellites equipped with QMs provide three orders of magnitude faster entanglement distribution rates than existing protocols where QMs are located in ground stations.  We analyse how entanglement distribution performance depends on memory characteristics, determine benchmarks to assess performance of different tasks, and propose various architectures for light-matter interfaces. Our work provides a practical roadmap to realise unconditionally secure quantum communications over global distances with current technologies.
%Global scale quantum networking with untrusted devices requires going beyond Quantum Key Distribution (QKD) with Weak Coherent Pulses (WCP) towards employing true single photons. However, the loss in optical fibres precludes any realistic application beyond few hundred kilometres. Quantum repeaters and satellite-based systems provide promising means to overcome this limitation. With this, current ambitious ground-based QR protocols remain limited to a few thousand kilometres and satellites have limited line-of-sight. In this article, we analyse the use of quantum memory (QM)-equipped satellites for quantum communication for quantum networking in the global scale. We focus on global range repeaters and Measurement-Device-Independent (MDI) QKD within the line-of-sight distance.  We demonstrate that satellites equipped with QMs provides three orders of magnitude faster entanglement distribution rates over global distances than existing protocols. For ground-based networks, we show that QMs can increase key rates for general line-of-sight distance QKD protocols. We determine meaningful benchmarks to the performance of memories for different tasks and propose different architectures for the light-matter interface. Our work provides a practical roadmap to realise unconditionally secure quantum communications over global distances with current technologies. 
\end{abstract}


\maketitle

\section{Introduction}
\label{S:1}

\noindent
Quantum technologies such as quantum computing~\cite{Deutsch1985, Shor1994}, communication~\cite{Bennett2014, Gisin2002} and sensing~\cite{Degen2017,Sidhu2020_AVS} offer improved performance or new capabilities over their classical counterparts. Networking, whether for distributed computation or sensing can greatly enhance their functionality and power. As one of the first applications of quantum communication, quantum key distribution (QKD) has been leading the emergence of quantum information technologies and establishes the foundation for wide-scale quantum networking~\cite{Lo2014_NP}. In QKD, the security of secret keys shared between two parties are guaranteed by the law of physics and not only through the computational power of an adversary. The last three decades have seen significant progress in QKD enabling technologies including hand-held devices ~\cite{8426754}, integrated optics fabrication~\cite{chip1}, and photon detectors~\cite{Boaron2018}.  

However, the main limitation to current implementations is the range over which a secure link can be established. Ground based QKD systems are inherently limited by in-fibre optical losses, specifically the key generation rate decreases exponentially with distance. By using cryogenically-cooled superconducting nanowire single-photon detectors (SNSPDs), Boaron \textit{et al.} have demonstrated secret key distribution of around 6 bit/s at a distance of 405 km~\cite{Boaron2018}. More recent \textit{twin-field} QKD~\cite{Lucamarini2018} methods have achieved $\sim$1 bit/hour over 404 km. Both of these demonstrations have utilised state of the art ultra-low loss optical fibres, with losses around 0.17 dB/km, such value being unlikely to improve significantly in a medium time horizon.

Conventional optical repeaters cannot be used with QKD as quantum information cannot be  deterministically cloned~\cite{Wootters1982,mazzascamp}, this is a reason for its security against eavesdropping. Current long-distance fibre QKD links employ ``trusted nodes'' that effectively relay a secure key between the end points. The trusted nodes are assumed to be safe from malicious parties and are potential points of weakness. Trusted nodes are also unsuitable for the long-range distribution of entanglement, hence the need to overcome the terrestrial limits ($\sim$ 1000 km) of direct quantum transmission.

Moving beyond these limits requires the use of intermediate nodes equipped with quantum memories (QM) or quantum repeaters (QRs) which do not need to be assumed to be free from malicious control (``untrusted''). By exploiting the assistance of QRs to divide the transmission link into smaller segments it is possible to overcome the fundamental rate-loss scaling for direct transmission, though at the expense of many intermediate repeater nodes (one every $< 100$~km) that could be costly and difficult to construct. Quantum repeaters perform local entanglement swapping operations to distribute entanglement across the whole link~\cite{Briegel1998, Duan2001, Sangouard2011, Muralidharan2016}. The use of repeater chains naturalise transmission links to arbitrary quantum networks that can be analysed and simulated using deep results from the classical network theory~\cite{Pirandola2019_CP}. Current fibre based QRs are still limited to around $\sim$4000 km \cite{Vinay2017} beyond which generation of meaningful keyrates (i.e. $\sim$1~Hz) becomes extremely challenging due to the need for large number of repeater stations. This falls short for a solution to global or intercontinental scale quantum communications.

The use of satellites may also extend QKD beyond the terrestrial direct transmission limit and is a natural approach to join different intercontinental fibre networks. Terrestrial free-space QKD is ultimately range limited by the Earth's curvature and the method is suitable mainly for intra- and inter-city links~\cite{Ursin2007,Sit2017}. In satellite QKD (SatQKD)~\cite{Liao2017, Yin2020}, the transmission loss through the vacuum of space is dominated by diffraction that has an inverse square scaling instead of exponential. However, the connection distance for SatQKD is primarily limited by the line-of-sight between satellite and ground station which in turn depends on its orbit unless the satellite acts as a trusted node~\cite{Liao2018,Oi2017,Sumeet2019, Vergoossen2020,quarc,scheduling}. To establish a global quantum network without trusted nodes will require overcoming the above limitations.  The use of quantum satellites equipped with quantum memories as quantum repeaters remains relatively unexplored.

In this paper, we develop and characterise new approaches for global QKD using space and ground networks. Our approach exploits satellites equipped with quantum memories to provide free-space optical repeater links to connect two end stations on the ground. We implement memory-assisted measurement-device independent QKD (MA-QKD) protocols~\cite{Abruzzo2014, Panayi2014, Luong2016} to achieve high rates and device-independent security on board satellites in a line-of-sight setting. The entanglement distribution rate is used as a benchmark to assess the performance of our repeater chain. Our approach overcomes limitations in purely ground-based repeater networks and trusted satellite relays to provide the best rate-loss scaling for quantum communications over planetary scales. Notably, we demonstrate that satellites equipped with QMs provides three orders of magnitude faster entanglement distribution rates over global distances than existing protocols. For connecting ground-based networks, we show that QMs can increase key rates for general line-of-sight distance QKD protocols. Our work provides a practical roadmap towards an implementation of global communication, navigation and positioning, and sensing. We conclude by providing meaningful benchmarks to the performance of QMs for different tasks and propose different architectures for the light-matter interface.

\section{Quantum repeater and memory-assisted QKD protocols}

\noindent
We first outline two QR protocols for global entanglement distribution followed by MA-QKD protocols in uplink and downlink configurations to increase the keyrates in a quantum communication within the line-of-sight distance of the satellite. Here, QMs are used as quantum storage devices to increase the rate of otherwise probabilistic Bell state measurements (BSMs) that form the backbone of most MDI protocols.

\subsection{Quantum repeaters}

\noindent
QRs can be grouped into different architectures depending on the error correction mechanism employed~\cite{Muralidharan2016}. The first generation of QRs rely on the postselection of entanglement, which acts as an entanglement distillation operation. Improved generations of QRs may employ active error correction codes that necessitate shorter link distances and higher number of qubits (50-100, i.e. a quantum computer in the Sycamore scale) per node. Hence, we restrict our attention to the first generation type architectures that employ ensemble-based QMs.
The use of atomic ensembles for long-distance communication was first proposed in a seminal paper by Duan, Lukin, Cirac and Zoller~\cite{Duan2001} also known as the DLCZ protocol. It relies on creating photon-spin wave entanglement through Raman scattering. This protocol has been modified and improved significantly over time~\cite{Sangouard2007, Chen2007, Sangouard2011}. Nevertheless, entanglement distribution rate with these schemes quickly drops below practically useful levels above few thousand kilometers which renders reaching true global distances a formidable challenge with land-based architectures. 

\begin{figure}
    \includegraphics[width =0.95\columnwidth]{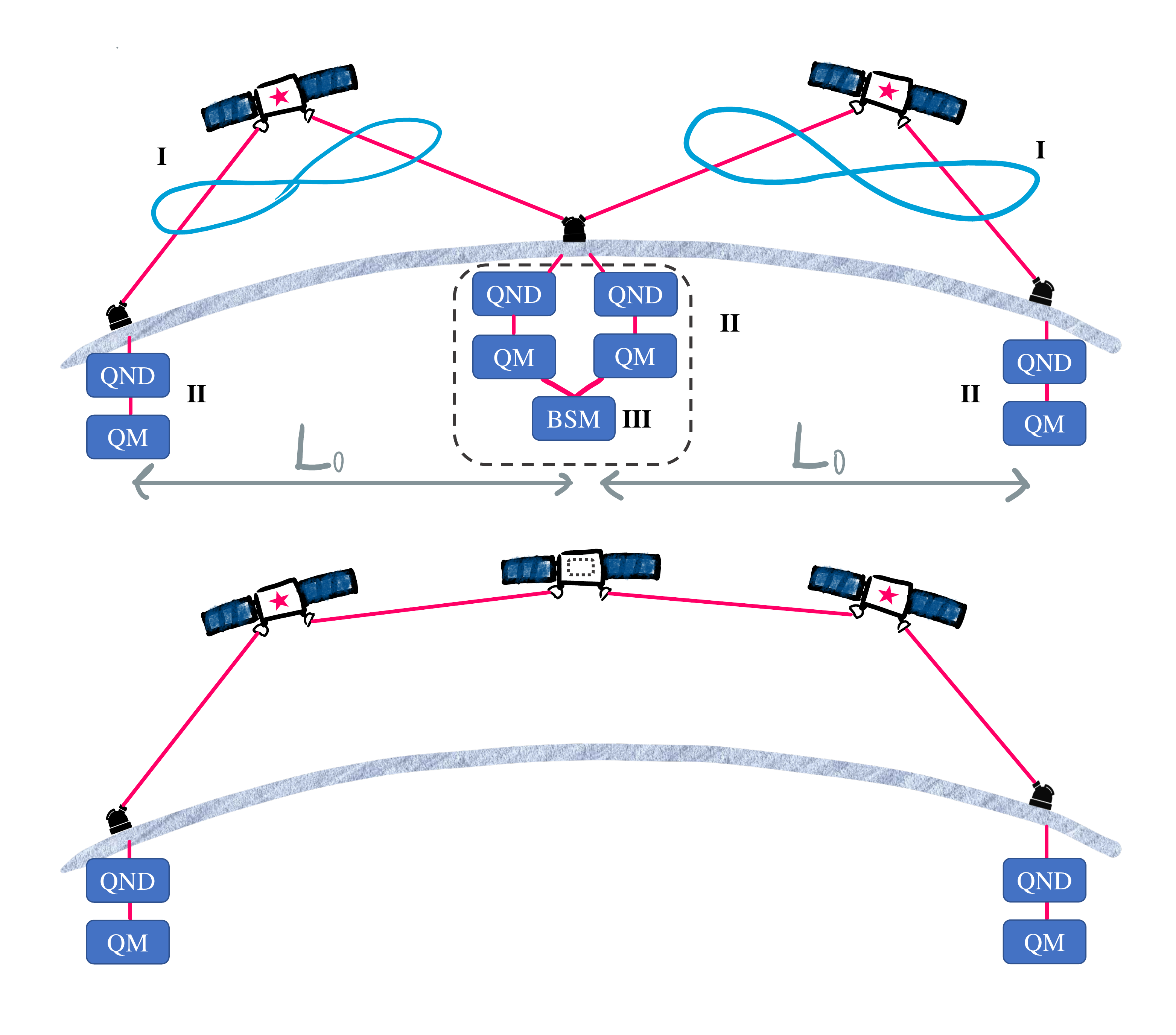}
    \caption{Top: Hybrid QND-QR protocol, following~\cite{Boone2014} with nesting level, $n=1$ and segment length, $L_0$. Entangled photon pairs are created by on-board sources (pink stars) and sent to ground stations (I). After a QND detection heralds the arrival of the photons they are loaded to QMs (II). BSM is performed between the memories to extend entanglement between end stations (III). Bottom: New architecture where the QND and QMs are also located on-board an orbiting satellite.}
    \label{fig:Boone}
\end{figure}

A hybrid, satellite-assisted architecture has been proposed for entanglement distribution with useful rates~\cite{Boone2014} (Fig.~\ref{fig:Boone}, top). It relies on satellites equipped with entangled photon pair sources communicating with the memory nodes located in ground stations. Other than the satellite links the main difference it exhibits with respect to other first generation protocols is that heralding is performed via a quantum non-demolition (QND) measurement. Entanglement is then distributed between the communicating parties via entanglement swapping operations between neighbouring nodes, similar to previous protocols. The authors cited technical challenges, such as launch and operation in space environment, in favour of locating the QMs in ground stations. However, during the 6 years since the proposal, atomic physics experiments have made a leap into space, mainly for atom interferometry and optical clock applications (outlined in Sec.~\ref{SecAtomSpace}). Thus the feasibility and performance of QR architectures that operate in space should be re-examined in light of these advancements (Fig.~\ref{fig:Boone}, bottom).

We consider a constellation with a total of $2^{n+1}-1$ satellites, where $n$ is the nesting level that divides the whole communication channel into $2^n$ segments. There are two types of satellites: one type carries a photon pair source and the other type QND and QM equipment for entanglement swapping (satellites with pink stars and the dashed box in Fig.~\ref{fig:Boone}). Such a  scheme will have several advantages over the original hybrid protocol. The first and most important is lower loss due to having only two atmospheric channels and the other inter-node links being located in space. The second advantage is that success will depend on the weather conditions only at two ground stations at the two ends of the communication link whereas the original proposal requires all ground stations (including intermediate relay stations) to simultaneously have good weather conditions, which becomes increasingly unlikely as the number of nodes increases. Finally the need for Doppler-shift compensation to ensure indistinguishability of photons in a BSM is greatly reduced due to lower relative inter-node velocities.

\subsection{MA-QKD schemes}
\label{sec:MAQKD}
\subsubsection{Uplink}
\begin{figure}
    \includegraphics[width =0.9\columnwidth]{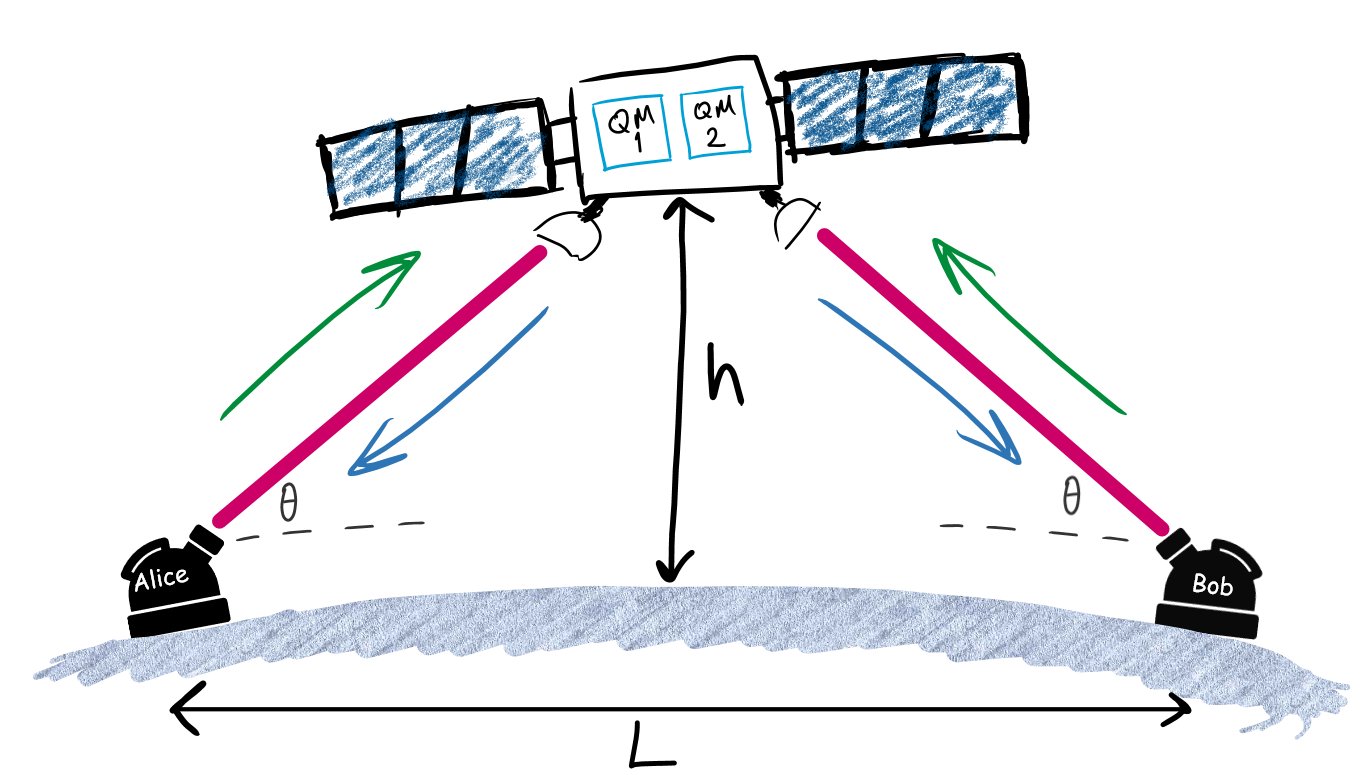}
    \caption{MA-QKD protocol, following~\cite{Panayi2014, Luong2016}, in the geometry of~\cite{Yin2017, Yin2020}. Green (blue) arrows show uplink (downlink) protocol. Alice and Bob both have standard BB84 encoders.
    QM: quantum memory, $\theta$: elevation, h: orbital height, L: total ground communication distance. }
    \label{figPanayi}
\end{figure}

\noindent
In a space-based setting, the protocol proposed in Refs.~\cite{Abruzzo2014, Panayi2014} relies on communicating parties on the ground using single photon sources with conventional BB84 encoders and sending them up to a satellite that acts as a middle station where they would each be stored in an individual QM. Memories will then be read out upon the successful heralded loading of both. A BSM is then performed on the retrieved photons to extract a key or perform entanglement swapping (Figure~\ref{figPanayi}). One of the key characteristics of this protocol is its high operating rate as there is no waiting time associated with the heralding signal travelling between the BSM station and the communicating parties. However, this geometry precludes the extension of this protocol into a repeater architecture. A central requirement of this scheme is the heralding of a successful memory loading process. Ref.~\cite{Panayi2014} analyses both direct and indirect heralding scenarios. The directly heralded scheme relies on the QND detection of incoming photons before being loaded into their respective QMs whereas the indirectly heralded scheme requires additional BSMs that herald the entanglement between the individual memories and the respective incoming photons. A BSM between the memories is then performed to distill a secret key. %\textit{Maybe talk about QND measurements for direct heralding?} 
The main drawback of the uplink geometry is the additional loss contribution from turbulence that can be as big as 20 dB compared with downlink transmission. Thus bigger receiver apertures in space are required which might be challenging to deploy.  %I would say that the main drawback with respect to the downlink case is the.....  

\subsubsection{Downlink}

\noindent
The other main MA-QKD protocol we analyze was first proposed in 2014~\cite{Luong2016}. Here the direction of travel of the photons is from the middle station to the communicating stations at the two ends. In this configuration each of the QMs in the middle station emit single photons that are entangled with the internal states of the respective memories towards the receiving ends. A BSM will be performed on the memories upon the successful BB84 measurements by the receiving parties. The repetition rate of the protocol is inherently limited by speed of light travel time of classical signals to herald a successful detection by the communicating parties to the middle station where the BSM is performed. This also requires long-lived QMs with storage times in the order of seconds to achieve similar performance to the previous method. An extension of this protocol in which the central pair of QMs are replaced with $m$ pairs of QMs~\cite{Trenyi2020} reduces the required storage time. 

\section{Results}

\noindent
In this section, we present numerical results for achievable entanglement (key) distribution times using space-based MA-QKD and QR architectures. We compare these results with known results that use ground-based and hybrid schemes. Calculation of the channel losses are given in Appendix~\ref{ChannelModel}.

\subsection{Space-based repeater calculations}
\label{RepCalcSec}

\noindent
In this section we analyse and compare two QR architectures, DLCZ and QND-QR in a space-based setting. 

The time required to create and distribute an entangled state with the DLCZ protocol is given by~\cite{Sangouard2011} 
\begin{equation}
T_{\mathrm{tot}}^{\mathrm{DLCZ}}=3^{n+1} \frac{L_{0}}{c} \frac{\prod_{k=1}^{n}\left(2^{k}-\left(2^{k}-1\right) \eta_m\eta_d\right)}{\eta_{d} \eta_{t} p (\eta_m\eta_d)^{n+2}} \label{eq:DLCZ}
\end{equation}
where we recall that $n$ is the nesting level that divides the whole communication length $L$ into $2^n$ links with $L_0$ length. $\eta_d$, $\eta_t$ and $\eta_m$ are detection efficiency, channel transmission and memory efficiency, respectively. We define $\eta_m = \eta_r\times\eta_w$ with $\eta_r$ ($\eta_w$) being the memory read-out (write) efficiency. Lastly, $p$ is the photon pair creation probability and $c$ is the speed of light. Memories should be pumped to create a sufficient rate of photon pairs, i.e. high $p$ but still low enough to minimise double pair emissions that scale as $p^2$. This assumes a single-mode memory and thus could be further reduced by using temporally multi-mode memories~\cite{Kutluer2017, Heller2020}. 

\begin{figure}[t!]
\begin{center}
\subfloat[Entanglement distribution time with total communication distance.]{\includegraphics[width=0.9\columnwidth]{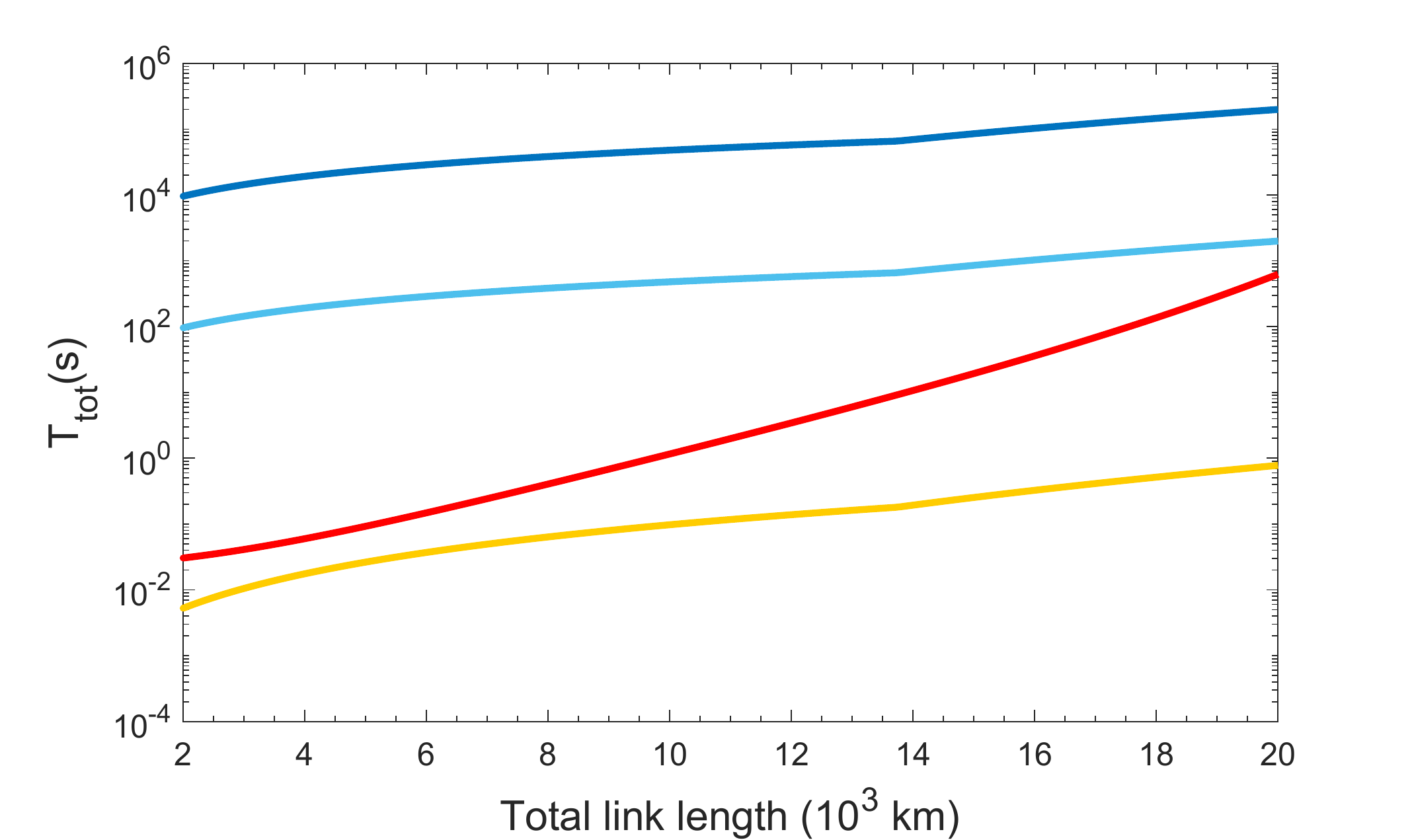}\label{fig:DistvsRate}} \\
\subfloat[Entanglement distribution time with beam divergence.]{\includegraphics[width=0.9\columnwidth]{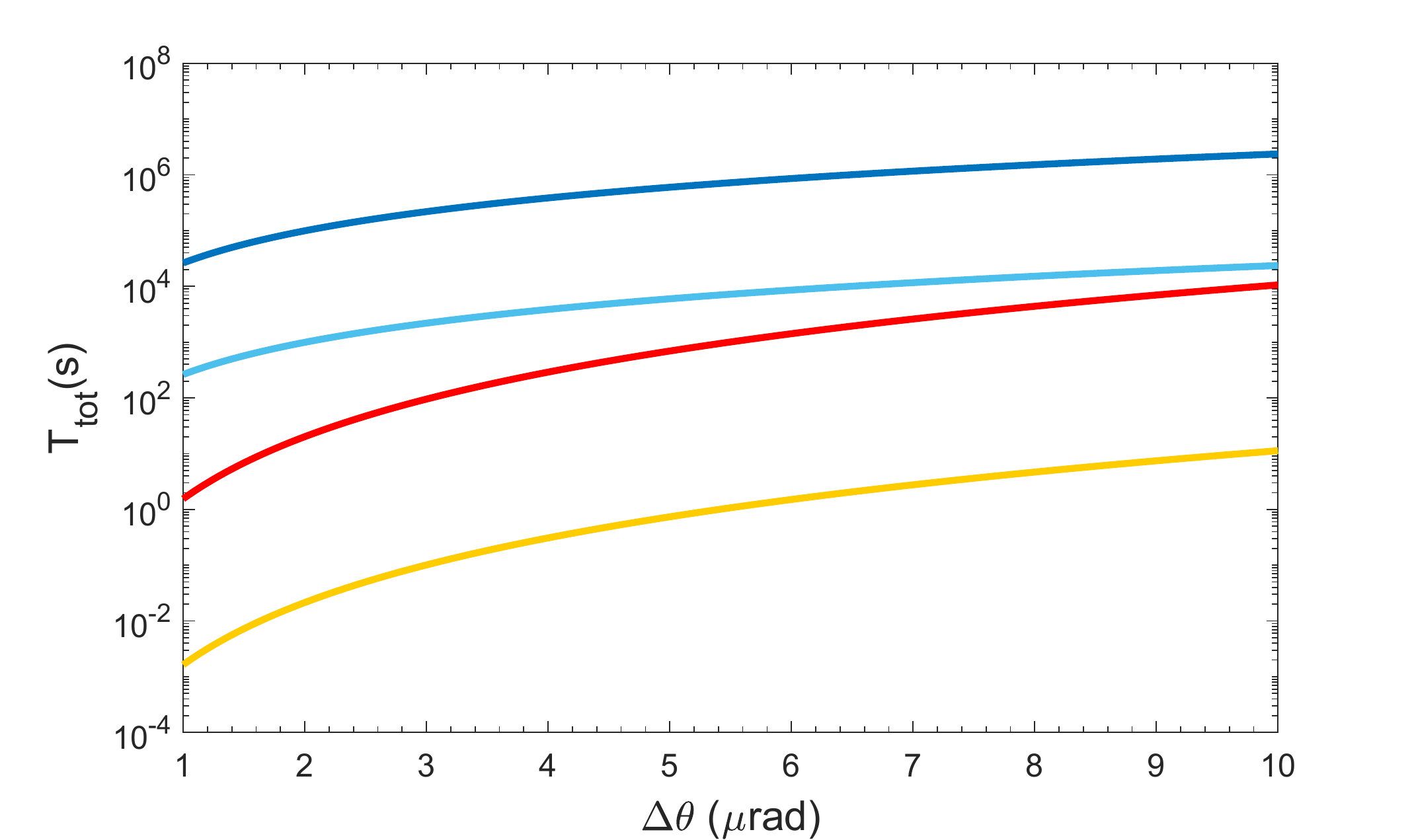} \label{fig:DivvsRate}}\\
\subfloat[Entanglement distribution time with memory efficiency.]{\includegraphics[width=0.9\columnwidth]{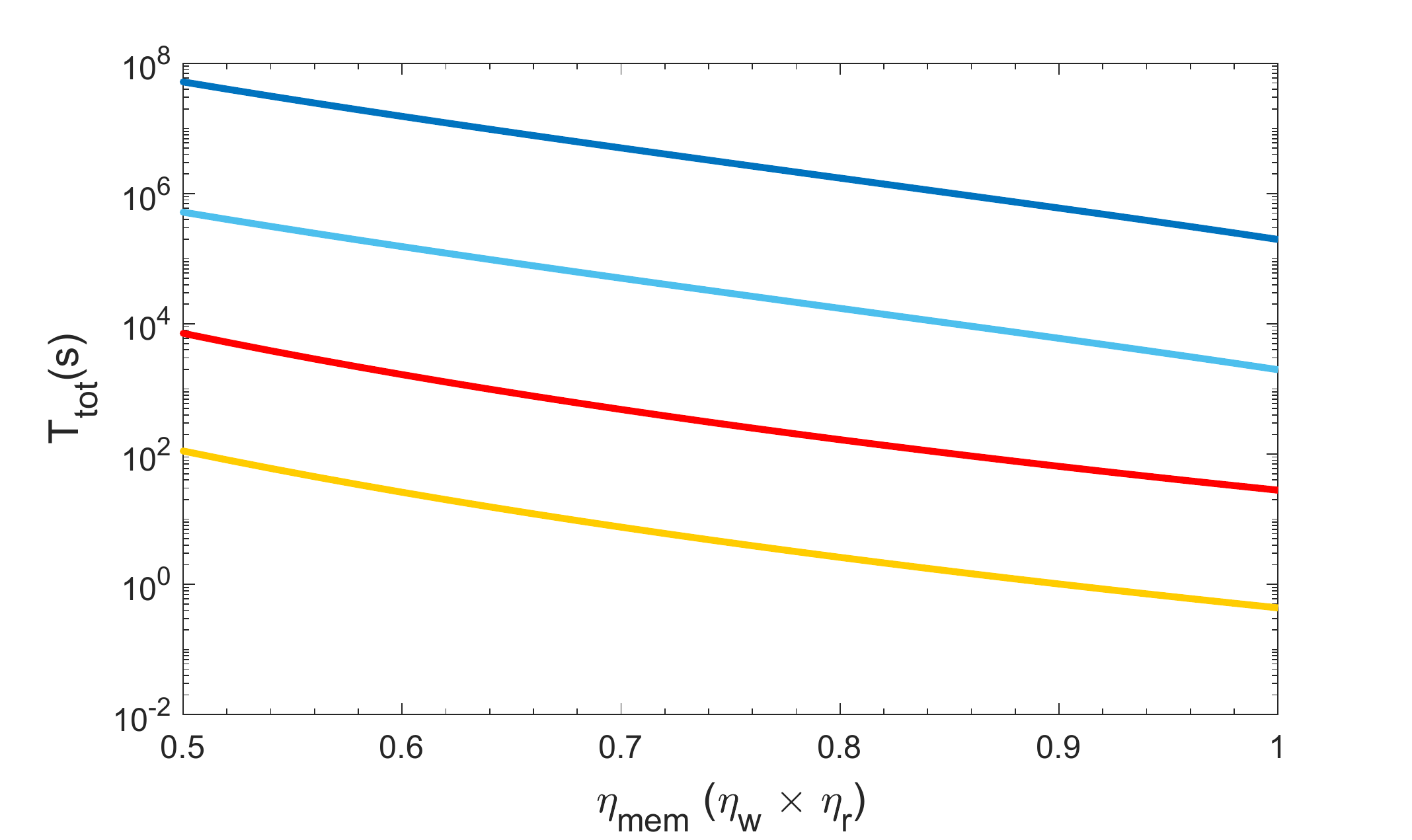} \label{fig:MemvsRate}}
\caption{Time to distribute an entangled pair as a function of (Fig.~\ref{fig:DistvsRate}) total distance, (Fig.~\ref{fig:DivvsRate}) beam divergence and (Fig.~\ref{fig:MemvsRate}) memory efficiency.  Within each plot: DLCZ with single (dark blue) or 100 mode (light blue) memory, hybrid-QND (red) and space-QND protocols (orange). The nominal assumed parameters (when not varied) are non-ideal Gaussian beams with divergence $\Delta\theta = 5~\mu$rad , $\mathrm{L} = 20000$~km and $\eta_{\mathrm{r}}\times\eta_{\mathrm{w}}\equiv\eta_{\mathrm{mem}} = 0.9$, with $\eta_{\mathrm{r}} = \eta_{\mathrm{w}}$. We fix $\eta_{\mathrm{q}} = 0.5$, $\eta_{\mathrm{s}} = 1$, $\eta_{\mathrm{d}} = 0.9$ and $R_\mathrm{s} = 20~\mathrm{MHz}$.}
\label{fig:Ttot_three}
\end{center}
\end{figure}%

On the other hand, entanglement distribution time in the QND-QR protocol is given by~\cite{Boone2014}
\begin{equation}
T_{\mathrm{tot}}^{\mathrm{QND}} = \left[R_{\mathrm{s}} \eta_{\mathrm{s}} P_{0}^{\mathrm{avg}} \eta_{\mathrm{q}}^{2} \eta_{\mathrm{w}}^{2}\left(\frac{2}{3} \frac{\eta_{\mathrm{r}}^{2}\eta_{\mathrm{d}}^{2}}{2}\right)^{n}\right]^{-1}.
\end{equation}
Here in addition to the parameters defined above, $\eta_q$ is the QND detection efficiency, $R_{\mathrm{s}}$ is the source repetition rate and $P_{0}^{\mathrm{avg}}$ is the average two-photon transmission.

The main difference between calculations presented in this section and in the original hybrid satellite-ground architecture is that only the two end links are satellite-ground links whereas all other optical channels are inter-satellite links. In Fig.~\ref{fig:Ttot_three} we present entanglement distribution times $T_{\mathrm{tot}}^\mathrm{DLCZ}$ and $T_{\mathrm{tot}}^\mathrm{QND}$. We assume a nesting level of $n=3$ in what follows. 
%We assume imperfect Gaussian beams with divergence $\Delta\theta = 5~\mu$rad, $\mathrm{L} = 20000$~km and $\eta_{\mathrm{r}}\times\eta_{\mathrm{w}}\equiv\eta_{\mathrm{mem}} = 0.9$, with $\eta_{\mathrm{r}} = \eta_{\mathrm{w}}$, when not scanned. We fix $\eta_{\mathrm{q}} = 0.9$, $\eta_{\mathrm{s}} = 1$, $R_\mathrm{s} = 20~\mathrm{MHz}$.

Entanglement distribution time as a function of total ground distance is plotted in Fig.~\ref{fig:DistvsRate}. DLCZ protocols are significantly slower than the QND protocols. The main reason is the long waiting times for the classical heralding signal transmitted between neighbouring nodes. It is expressed with the factor $L_0/c$ in Eq.~\ref{eq:DLCZ} and accumulates as the distance, hence the loss, increases. 
Hybrid ground-space and full-space QND protocols start off within an order of magnitude but the scaling quickly turns against the hybrid protocol as atmospheric loss increases due to the increasingly narrow grazing angle and dominates the diffractive loss. The Space-QND protocol offers 3 orders of magnitude faster entanglement generation rates for global distances.

In Fig.~\ref{fig:DivvsRate} we       plot $T_\mathrm{tot}$ as a function of beam divergence ($e^{-2}$ beam divergence half-angle, Eq.~\ref{eq:div}), $\Delta\theta$. Diffraction limited beams at optical wavelengths has around 1~$\mu$rad divergence for telescope radii of around 20~cm. QND protocols are more sensitive to channel losses since they scale with $\eta_{t}^{-2}$ whereas DLCZ schemes follow $\eta_{t}^{-1}$ scaling. This sensitivity results in $\sim$4 orders of magnitude slower operation times with an imperfect beam with 10~$\mu$rad divergence (similar to MICIUS) with respect to what can be achieved with a diffraction limited beam. The scaling difference between DLCZ and QND protocols results in hybrid-QND scheme having a comparable speed with the multimode DLCZ at large divergences. Although optical links (in the limit of large grazing angle) do not suffer from exponential losses such as in optical fibers, this example shows it is nevertheless crucial to have high quality beams with very small divergence.   

Lastly, we investigate the effect of the finite memory efficiency on the entanglement distribution time in Fig.~\ref{fig:MemvsRate}. We again see that it is highly crucial to have highly efficient memories. For QND protocols, 50\% memory efficiency reduces the operation speed by more than two orders of magnitude when compared to 90\% memory efficiencies. Given that satellites only have few minutes of flyby over any target, this difference easily makes the whole protocol impractical. 

$T_{\mathrm{tot}}$ dictates the minimum required storage time for the QM used in the repeater chain. If we look at Fig.~\ref{fig:DistvsRate}, the full space-based protocol proposed here requires a storage time of around 100~ms for $10^4$~km ground distance and a 700~ms storage is required for a distance half the Earth's circumference. On the other hand, the hybrid protocol necessitates $\sim1.2$~s and $~9.5$~mins for the same distances.

So far we have concentrated on global scale quantum networking via satellite links. In what follows we will analyse the MA-QKD proposals in a shorter range, line-of-sight setting.

\subsection{Results for MA-QKD protocols}

\begin{figure}
    \includegraphics[width =1\columnwidth]{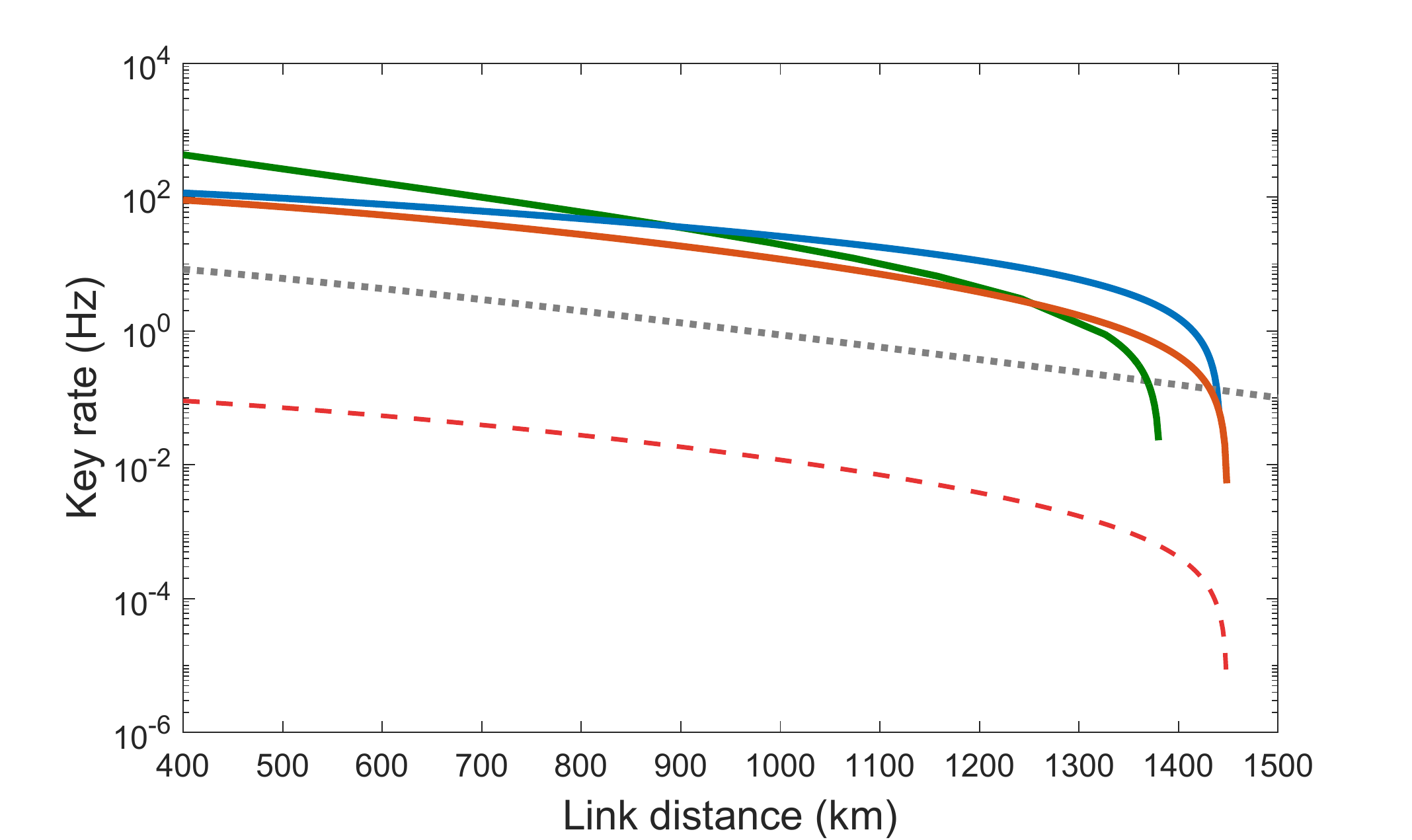}
    \caption{Key rates with E91 protocol, i.e. no memory and MA-QKD schemes. Gray dotted: E91 ($R_s = 20$~MHz); blue: uplink configuration with storage time 5 ms; solid (dotted) red: downlink configuration, with N=1000 (single) temporal modes with storage time 7.5~s; green: downlink with $N=1000$ temporal modes and $m=100$ memory pairs with storage time 100~ms.}
    \label{CompareAll}
\end{figure}

\noindent
For shorter, continental distances one solution is to eliminate the need for QMs by increasing the brightness of the photon sources. However,  GHz-rate entangled photon pair generation remains a challenging task~\cite{Zhang2015}. Even with deployment of such fast sources, GHz-bandwidth QMs~\cite{Guo2019} would be still useful to increase the achieved key rates. 
In this section we adapt the well-established MA-QKD protocols explained in Sec.~\ref{sec:MAQKD} to a space-based scenario. We will benchmark the calculated key rates with MA-QKD protocols against the well-known QKD protocol with entangled photons~\cite{Ekert1991, Ma2007}, known as E91. Figure \ref{CompareAll} shows the achievable key rate as a function of ground distance $L$ with i) E91 protocol, i.e. no QM (gray dashed); ii) uplink configuration with protocol presented in Ref.~\cite{Abruzzo2014, Panayi2014} (blue) and finally iii) downlink scenario with $m=1$ (red) and $m=100$ (green)~\cite{Trenyi2020}. Parameters used in simulations to generate Fig.~\ref{CompareAll} are shown in Table~\ref{table:QKD}.

\begin{table}[ht]

\centering
\begin{tabular}[t]{lccc}
\hline
\textbf{Description}&\textbf{Parameter} &\textbf{Downlink}  & \textbf{Uplink}\\
\hline
Orbital height& $h$ &400~km &400~km\\
Sender aperture radii\hspace{1cm}& R$_{\text{sender}}$&15 cm &15 cm\\
Receiver aperture radii& R$_{\text{receiver}}$&50 cm&50 cm\\
Divergence& $\Delta\theta$&$10~\mu$rad&$10~\mu$rad\\
Storage time& $\tau$ &100~ms, 7.5~s&5~ms\\
Memory pairs& $m$ &100, 1&1\\
%R$_\text{rep}$&$L_0/c$&10~MHz\\
Memory efficiency& $\eta_{\text{mem}}$&0.8&0.8\\
Detector efficiency& $\eta_{\text{det}}$&0.7&0.7\\

\hline
\end{tabular}
\caption{Parameters used in MA-QKD simulations.}
\label{table:QKD}
\end{table}%

\noindent
The model presented here predicts a secret key rate of 0.15~bits/s at 1120~km ground distance with 5.9~MHz repetition rate without memories and this value is consistent with the recently reported value 0.12~bits/s by the MICIUS team~\cite{Yin2020}. For the simulations discussed here, we assume a repetition rate of 20~MHz that yields around 1~bit/s at 1000~km ground distance. We will use this value to benchmark the performance of MA-QKD schemes in the following sections.

%--

\subsubsection{Uplink scenario}
\noindent
The blue curve in Fig.~\ref{CompareAll} shows the expected key rate obtained with the protocol in Ref.~\cite{Panayi2014}. We assume 15~cm (50~cm) of radius for sender (receiver) telescope, with $10~\mu$rad beam divergence and we omitted the atmospheric turbulence. The memory is assumed to perform with a storage time of 5~ms and $80\%$ combined write-read efficiency. Operation rate is assumed to be 20 MHz and we only consider single-mode memory case as the operation rate is not limited to any classical communication between parties. As can be seen, this protocol offers a speed up over the no-memory protocol up to $\sim1450$~km after which no key could be generated. 

\begin{figure}[t!]
\begin{center}
\subfloat[Uplink configuration with $m=1$.]{\includegraphics[width=1\columnwidth]{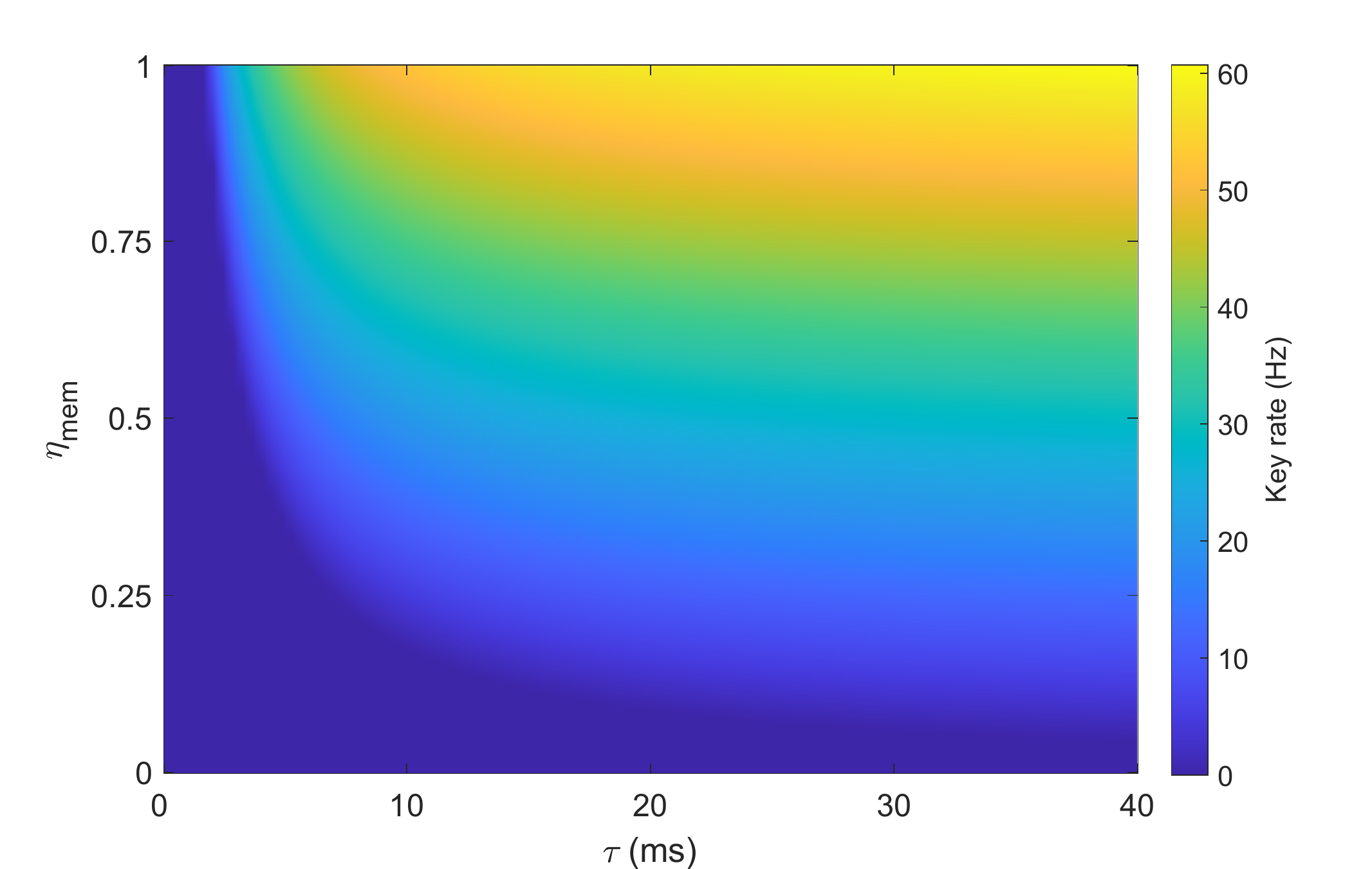} \label{fig:2DPanayi}}\\
\subfloat[Downlink configuration with $m=1, N=1000$,.]{\includegraphics[width=1\columnwidth]{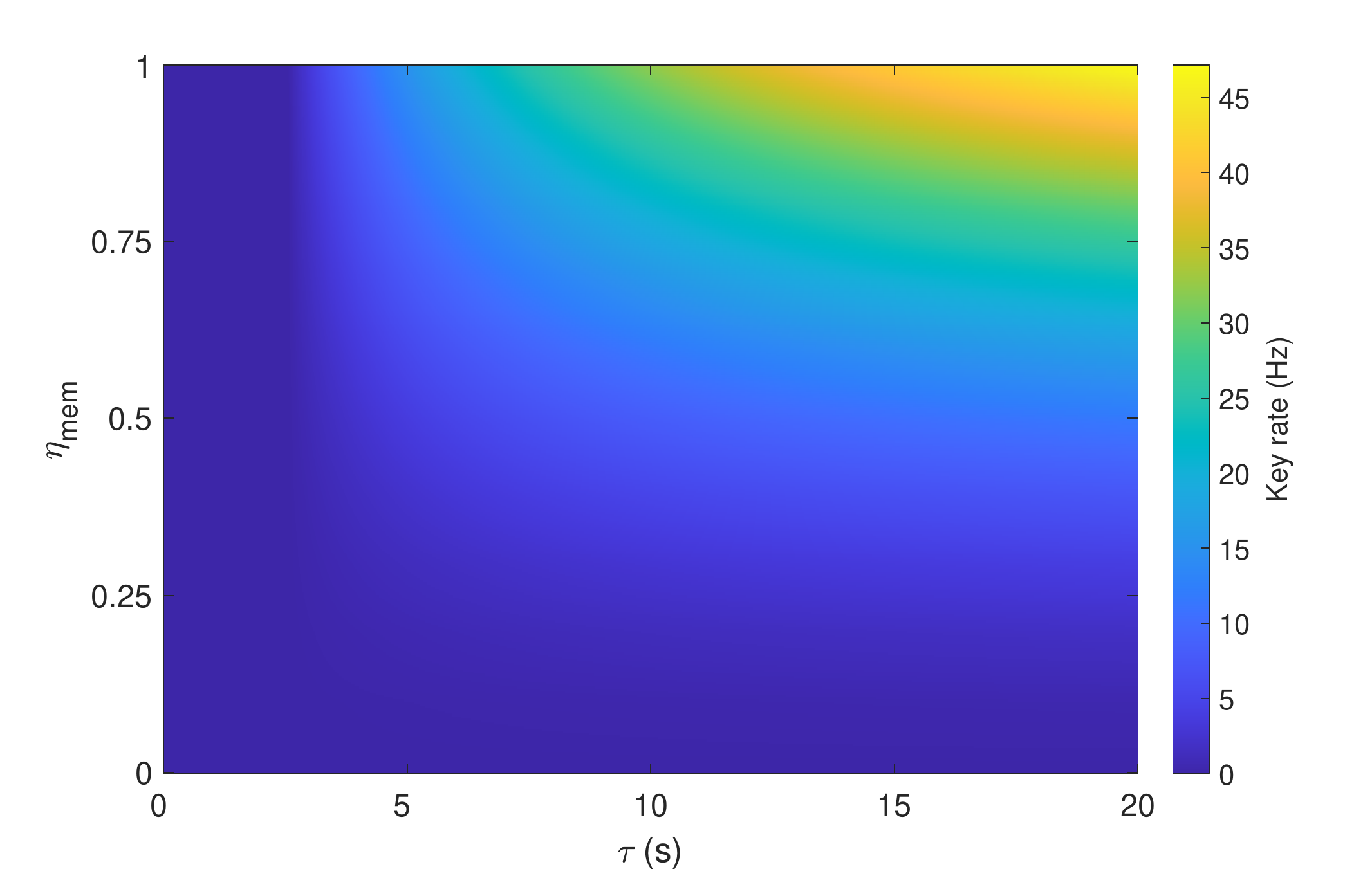}\label{fig:2DLuong}} \\
\subfloat[Downlink configuration with $m=100, N=1000$.]{\includegraphics[width=1\columnwidth]{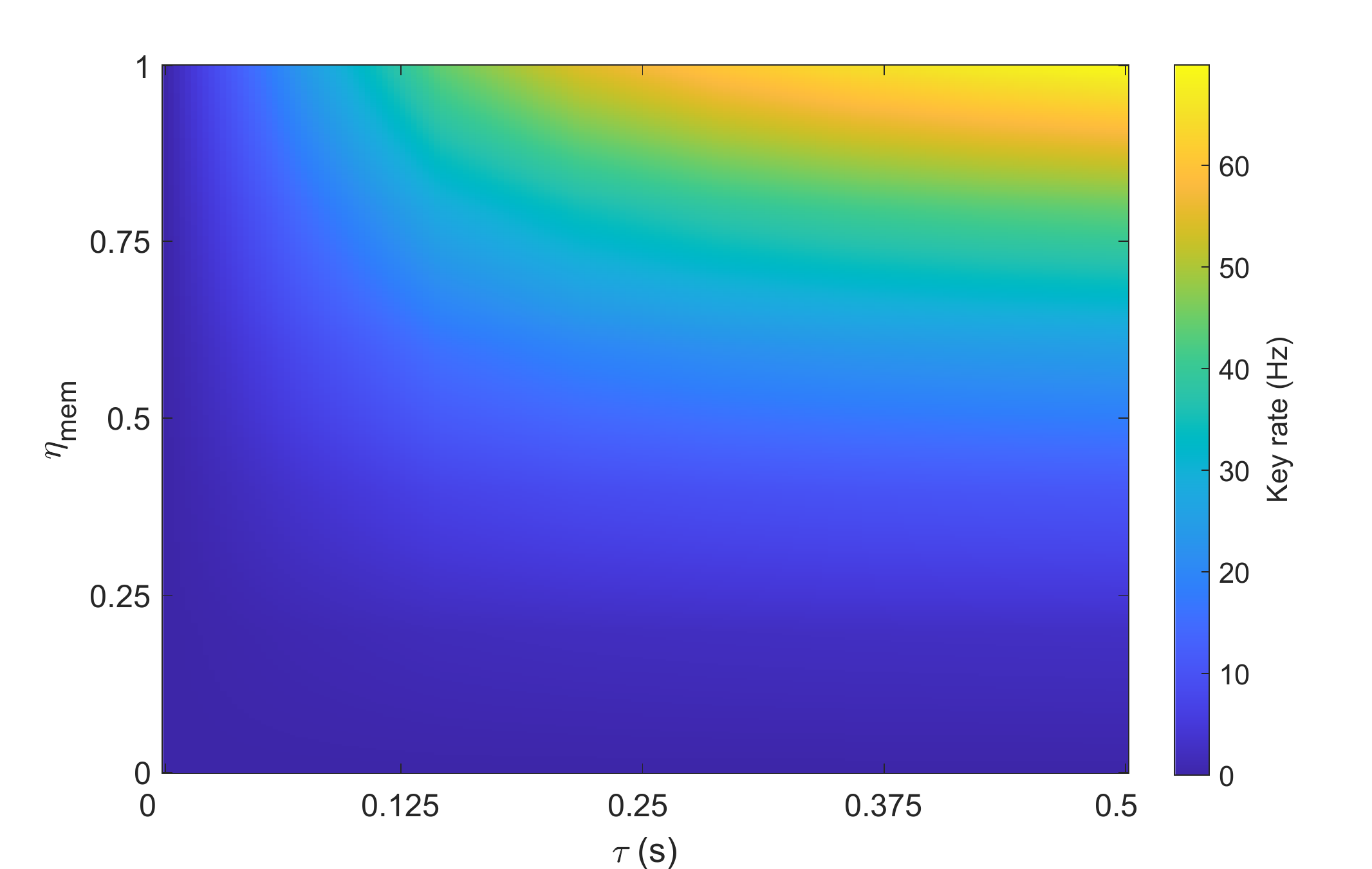} \label{fig:2DLuongMM}}\\
\caption{Achievable key rates at 1000~km ground distance as a function of memory time and efficiency for different configurations.}
\label{fig:QKD_three}
\end{center}
\end{figure}

Figure \ref{fig:2DPanayi} shows the achievable key rate at a fixed $L_0 = 1000$ km as a function of memory write efficiency and storage time.It is observed that no meaningful key rate could be achieved with memory dephasing times of $<5$ ms, regardless of the memory efficiency. The dependence on the memory efficiency is less dramatic for the uplink protocol. With a relatively modest storage time of 20 ms and a memory write efficiency of around 50\%, one can achieve more than an order of magnitude improvement over the no-memory case as summarized above. However, one should note that this architecture can not be extended to a quantum repeater architecture due to the photon travel direction precluding any entanglement swapping operation between neighbouring links. 

\subsubsection{Downlink scenario}

\noindent
The operation rate of the downlink protocol is intrinsically limited (for a single-mode memory) by  the time the classical signal takes to reach the other party, i.e. $R = c/2L_{\mathrm{LoS}}$, with $c$ being the speed of light and $L_\mathrm{LoS}$ is the line-of-sight distance between a ground station and the satellite. Hence, this protocol requires long-storage times, in the order of seconds. The parameters are shown in Table~\ref{table:QKD}. In Fig.~\ref{CompareAll} the dotted red curve shows the key rate with a single-mode QM. The achievable key rate is significantly lower than the E91 protocol due to the slower operation speed. The only way to increase the key rate is thus to operate with temporally multimode QMs. The solid red curve shows the key rate that is only possible with a QM that could store 1000 temporal modes. This provides an enhancement of around an order of magnitude between line-of-sight distances of 500-1000 km.
Figure~\ref{fig:2DLuong} shows the achievable key rate as a function of the memory efficiency and dephasing time at a fixed ground distance of 1000 km, with N=1000 temporal mode QM. At such a distance storage times shorter than 5 s would not be sufficient for the protocol to produce any meaningful key rate regardless of the storage efficiency. Likewise, storage efficiency of around 35\% is needed in combination with a $\tau = 10$~s to reach a 10~Hz key rate. 

We further analysed the extension of this protocol with $m=100$ pairs of QMs located in the middle station. Green curve in Fig.~\ref{CompareAll} shows that storage time of only 100~ms is sufficient instead of the very demanding 7.5~s to reach the same distance with similar keyrates. Fig.~\ref{fig:2DLuongMM} shows the performance map of this scheme again at a fixed ground distance of 1000~km. The striking feature here is that the cut-off storage time below which no key could be transmitted is only a few ms. 

With these findings in mind, we discuss potential memory types suitable for space missions in the next section.

\section{Towards first possible quantum memory experiments in space}

\noindent
In this section we will overview the existing QM experiments and provide a roadmap towards choosing a proper physical system in light of the findings of the previous section. We focus on ensemble based systems as it would be more straightforward to implement temporally multimode storage needed in the protocols described in this paper (see Appendix~\ref{sec:TempMM}). However, we note that the first land-based MA-QKD experiment has been recently performed with a single color center in diamond at $\sim$mK temperatures~\cite{Bhaskar2020}.

\subsection{Warm vapour systems}

\noindent
Photon storage in the long-lived ground states of alkaline vapors at room temperature is particularly appealing, as it requires neither complex cooling mechanisms nor large magnetic fields. This makes such memories ideal for field applications in remote environments, e.g. under sea or in space. The  performance of warm vapor memories has been continuously improved since the first demonstrations of memories based on electromagnetically induced transparency (EIT) in the 2000s. In recent years, the development of quantum memory implementations in alkaline vapor have gained remarkable momentum: (1) A vapor cell memory reached a storage time of $\tau$ = 1~s by using spin-orientation degrees of freedom and anti-relaxation coatings
~\cite{Katz2018}. (2) The efficiency of a room temperature EIT-like memory was pushed beyond 80\%~\cite{Guo2019}.  (3) EIT-like quantum memories with $\sim1$~GHz bandwidths were developed~\cite{Wolters2017}. These could in principle be extended to the storage of multiple signals in individually addressable sub-cells, as realized in cold atomic ensembles~\cite{Pu2017}. 
 
Besides ground state EIT memory, another promising vapor cell memory concept is the storage of photonic quantum information in highly excited atomic orbitals. These orbitals are relatively long lived, allowing for storage times on the order of 100 ns. The fast ladder memory scheme is based on two-photon off-resonant cascaded absorption~\cite{Kaczmarek2018, Finkelstein2018}. This scheme allows for the virtually noise-free storage with acceptance bandwidths in the GHz regime, but it needs to be further developed to allow for the comparable long storage times required by long-distance quantum communications.

\subsection{Laser-cooled atomic ensembles}\label{SecAtomSpace}

\noindent
Laser-cooled atomic systems are well-established platforms for quantum information storage. High efficiency~\cite{Bao2012}, temporal~\cite{Heller2020} and spatial multimode storage~\cite{Pu2017} have been performed among many other experiments in the last years. There has been a growing interest in deploying cold-atom experiments in space for more than a decade. This is driven by a combination of desire for access to longer periods of microgravity for fundamental research, and the deployment of instruments such as optical clocks on satellites for future global positioning concepts. Cold atom ensembles and Bose-Einstein condensates have already been created on orbiting platforms including %microgravity platforms such as parabolic flights~\cite{Langlois2018}, drop-towers~\cite{Muntinga2013}, sounding rockets~\cite{Becker2018}, and now even on 
Tiangong-2~\cite{Liu2018} and the ISS~\cite{Elliott2018, Aveline2020}. %Further launches to ISS are planned such as the ACES/PHARAO atomic clock~\cite{Laurent2015}, or the dual-species atomic experiment facility BECCAL~\cite{Frye2019}.  Various proposals are in development for deployment on cubesats such as CASPA~\cite{Bongs2019} or as part of larger proposals such as AEDGE~\cite{ElNeaj2020}. 

\subsubsection{BECCAL: Cold atom experiment on board ISS}

\noindent
In the context of these platforms, BECCAL is of particular interest due to the variety of experiments it is designed to perform. These experiments include the possibility of conducting initial demonstrations of optical memories in space. In short, BECCAL~\cite{BECCAL2019} will be capable of producing 3D-MOTs of \num{2e9} $^{87}$Rb atoms, \num{1e9} $^{85}$Rb atoms, \num{8e8} $^{39}$K atoms, \num{4e8} $^{41}$K atoms, and \num{1e7} $^{40}$K atoms in single species operation, it will also be possible to obtain single species BECs of \num{1e6} $^{87}$Rb atoms, or \num{1e5} $^{41}$K atoms. Atoms can also be confined in a \SI{1064}{\nano\meter} dipole trap with a waist of \SI{100}{\micro \meter} and a tunable potential depth of \SIrange{0.01}{5}{\micro \kelvin}.  Quantum coherences of longer than \SI{5}{\second} are planned.  Due to the absence of gravity, atomic samples can be used without additional, gravity compensating, trapping potentials.
Within BECCAL, the possibility for optical memory experiments is mediated via the detection scheme. Absorption detection is performed via two perpendicular axes to allow the gathering of three dimensional information about atom clouds. Via a distribution and switching system, it is possible to deliver light addressing the D2-lines of rubidium and potassium in a variety of pulse schemes (the $5^2S_{1/2} \rightarrow 5^2P_{3/2}$ and $4^2S_{1/2} \rightarrow 4^2P_{3/2}$ transitions in Rb and K respectively). One can deliver `cooling' and `repump' frequencies (i.e. F=2 $\rightarrow$ F'=3 and F=1 $\rightarrow$ F'=2 respectively for ${}^{87}$Rb) simultaneously or consecutively on a single axis, or in a crossed beam arrangement with cooling on one axis and repump on the other. Each frequency can be switched independently in less than \SI{1}{\micro\second}. These flexible conditions will facilitate storage techniques such as EIT in a microgravity environment thus being a pathfinder and demonstrator for the technology discussed in this paper.
%Thanks to the microgravity environment, the atomic clouds can be freely floating. Additionally, the densities could be increased by trapping the atoms in a single or crossed dipole trap at 1064~nm (trap frequencies? from paper?). 
%In this configuration, the two detection arms can simultaneously deliver cooling and repump light for either or both rubidium and potassium.  The two imaging axes can also be used to deliver 
%When used for imaging the light on each axis is identical, though the axes can be extinguished individually These two detection axes are also capab
%and the possibility of independent application, switching and control of cooling and repumping light on at least two orthogonal axes. Thus facilitating the storage techniques such as EIT in a microgravity environment as a pathfinder and demonstrator for the technology discussed in this paper.

\subsection{Rare-earth ion doped crystals (REIDs)}

\noindent
REIDs are another major platform to realize QMs in the context of quantum communication. Some of the recent advances include but are not limited to; demonstration of quantum correlations between long-lived hyperfine states and telecom photons~\cite{Seri2017}, storage of bright pulses up to a minute long duration~\cite{Heinze2013} and demonstrating 6-hour coherence time~\cite{Zhong2015}. The other research front in REID field is the miniaturization of these experiments. Waveguide geometries~\cite{Marzban2015,Corrielli2016} offer an enhanced compactness. The storage bandwidth is usually limited to a few MHz due to narrow hyperfine level separation however recent electronic-nuclear hybrid storage protocols would open up possibilities of storing large bandwidth photons in the long-lived spin states~\cite{Businger2020}, this would enable higher operation rates, $R_s$. Combination of compactness, high-bandwidth storage capability together with high efficiency and long storage times would place REID systems at the forefront of QM systems for space applications.  
On top of material considerations, REIDs are also suitable for temporally multimode storage~\cite{Gundogan2015, Jobez2016, Kutluer2017}. REIDs could be a promising candidate to realize a space-based QR With the development of miniature, space-compatible cryostats~\cite{You2018}.

\section{Conclusions}

\noindent
Quantum cryptography is the framework behind novel entanglement distribution protocols and security proofs. It has rapidly developed from simple lab demonstrations to in-field applications. However, developing and implementing robust QKD protocols over global transmission lengths remains an open challenge. The use of both ground and satellite based quantum repeater networks provide the most promising solution to extend quantum communications to global scales.

In this work, we provide the first theoretical analysis towards this goal~\footnote{After completion of this work we were made aware of a recent new preprint that also explored the use of quantum memories in space~\cite{Liorni2020}. Our study provides a more complete investigation of the use of quantum memories on key rates for concrete line-of-sight distance QKD protocols. We also analyse the effect of different experimental parameters such as beam divergence and memory efficiency on the performance of these protocols. On the other hand \cite{Liorni2020} includes a cost analysis and considers different orbital configurations.}. Our proposal uses satellites equipped with quantum memories in low Earth orbit that implement measurement-device independent QKD. We benchmark entanglement distribution times achieved through our architecture with existing protocols to find an improvement of $\sim$3 orders of magnitude over global scales. With the majority of optical links now in space, a major strength of our scheme is its increased robustness against atmospheric losses. We demonstrate that quantum memories can enhance secret key rates in general line-of-sight QKD protocols using practical devices that would be available in the near future. Our work provides a practical roadmap towards the implementation of quantum memories for space based fundamental physics experiments and paves the way to a promising realisation of a global networked quantum communications infrastructure.

%Our work leads to further research questions. It would be interesting to explore the effects of finite block sizes on the key rate. We anticipate a pronounced effect owing to a transmission time window between satellites and ground stations. 

Our work leads to further interesting research questions. It would be interesting to explore the effects of different satellite orbits (geostationary and medium Earth orbit), orbital dynamics, and constellation designs on entanglement distribution times. This naturally leads to the question of engineering efficient satellites network topologies, where quantum memories with even modest coherence times can effect profound gains to entanglement rates~\cite{Pant2019_NPJ}. Moreover, it would be interesting to explore the practical effects of finite block sizes on the key rate~\cite{Lorenzo2020}, for example effects owing to a transmission time window between satellites and ground stations.

\section{Acknowledgements}

\noindent
MG and MK acknowledge the support by the German Space Agency DLR with funds provided by the Federal Ministry of Economics and Technology (BMWi) under grant numbers 50WM1958 (OPTIMO), 50WM2055 (OPTIMO-2), which made this work possible. VH acknowledge the support by the German Space Agency DLR with funds provided by the Federal Ministry of Economics and Technology (BMWi) under grant number 50WP1702 (BECCAL).

DKLO is supported by the EPSRC Researcher in Residence programme (EP/T517288/1). JSS, LM, and DKLO acknowledge the travel support by the EU COST action QTSpace (CA15220), and LM acknowledges the travel support by Scottish Universities Physics Alliance SUPA (LC17683). Funding was provided by the UK Space Agency through the National Space Technology Programme (NSTP3-FT-063 ``Quantum Research CubeSat'', NSTP Fast Track ``System Integration \& Testing of a CubeSat WCP QKD Payload to TRL5''), EPSRC Quantum Technology Hub in Quantum Communication Partnership Resource (EP/M013472/1) and Phase 2 (EP/T001011/1), and Innovate UK (EP/S000364/1). 

MG acknowledges Guillermo Curr{\'a}s Lorenzo for double-checking the code for MA-QKD calculations and pointing to few minor errors at an early stage of the work.

%\section{Author Contributions}

%\noindent
%MG, JW, DKLO and MK conceived this project. MG designed the study and performed the calculations with inputs and feedback from all authors. MG and JSS wrote the manuscript with contributions from all authors. MK supervised the project.
%\section{Competing Interests}
%\noindent
%The authors declare no competing financial interests.

%\section{Materials and Correspondence}

%\noindent
% EPSRC Data Handling Statement required.
% This work is entirely theoretical, there is no data underpinning this publication.
%Correspondence and material requests should be addressed to Mustafa G\"{u}ndo\u{g}an (mustafa.guendogan@physik.hu-berlin.de).

\appendix

\section{}
\label{ChannelModel}
\subsection{Quantum link modelling and channel losses}

\noindent
An important requirement for the estimation of the performance of a space-based quantum communication systems is the precise modelling of the optical loss and source of noise of the channel as they both decreases the secret key rate. The former by making the transmitted quantum states less distinguishable by the receiver, and thus decreasing the overall detection rate, and the latter by increasing the Quantum Bit Error Rate (QBER). 

{\itshape Diffraction losses:} the dominant source of loss is diffraction which for a Gaussian mode of initial beam waist $\omega_0$ and wavelength $\lambda$ travelling a distance $d$ is given by \cite{gagliardi}:

\begin{equation}
    \eta_{\text{dif}}=1-\exp\left[-\frac{D^2_R}{2\omega_d^2}\right]\quad\mbox{with}\quad \omega_d^2=\omega^2_0\left[1+\Big(\frac{\lambda d}{\pi \omega_0^2}\Big)^2\right]
\end{equation}
Where $\omega_d$ is the accumulated beam waist at distance $d$ from the source and $D_R$ is the receiver aperture. As one can see, diffraction losses can be mitigated by increasing the receiver aperture but this could be unfeasible due to payload constraints. However, one should note that that diffraction losses scales quadratically with the link distance (for $d\gg \pi \omega_0^2/\lambda$) contrary to the exponential scaling for a fiber link with the length of the fibre. The divergence $\Delta\theta$ of an imperfect Gaussian beam is characterized by its $M^2$ value through the following relation: 
\begin{equation}\Delta\theta=M^{2} \frac{\lambda}{\pi \omega_{0}}.\label{eq:div}\end{equation}

{\itshape Atmospheric losses:}  Atmosphere constituents cause absorption and scattering of the optical signal, those effects depends on the signal wavelength. The atmospheric loss that includes absorption as function of elevation angle of $\theta$ is given by~\cite{Sumeet2019}:
\begin{equation}\eta_{\text{atm}}(\theta)=\left(\eta^{\pi/2}_{\text{atm}}\right)^{\csc \theta}\end{equation}
%\begin{equation}\chi_{\mathrm{ext}}(\theta)=\exp [-\beta \sec (\theta)]\label{eqtn:loss}\end{equation}
Here $\eta^{\pi/2}_{\text{atm}}$ is the transmissivity at Zenith and can be computed from a given model for the atmospheric absorption $\gamma(r;\lambda)$ as:
\begin{equation}
    \eta^{\pi/2}_{\text{atm}}=\int_0^h dr \; \gamma(r;\lambda)
\end{equation}
where $h$ is the altitude of the satellite. The value of $\eta^{\pi/2}_{\text{atm}}$ can also being found using dedicated software such as MODTRAN \cite{modtran}, at 780 nm the Zenith transmissivity is about 80$\%$. 

\begin{figure}[t]
\centering
\subfloat[Channel loss with satellite to ground distance.]{\includegraphics[width=0.9\columnwidth]{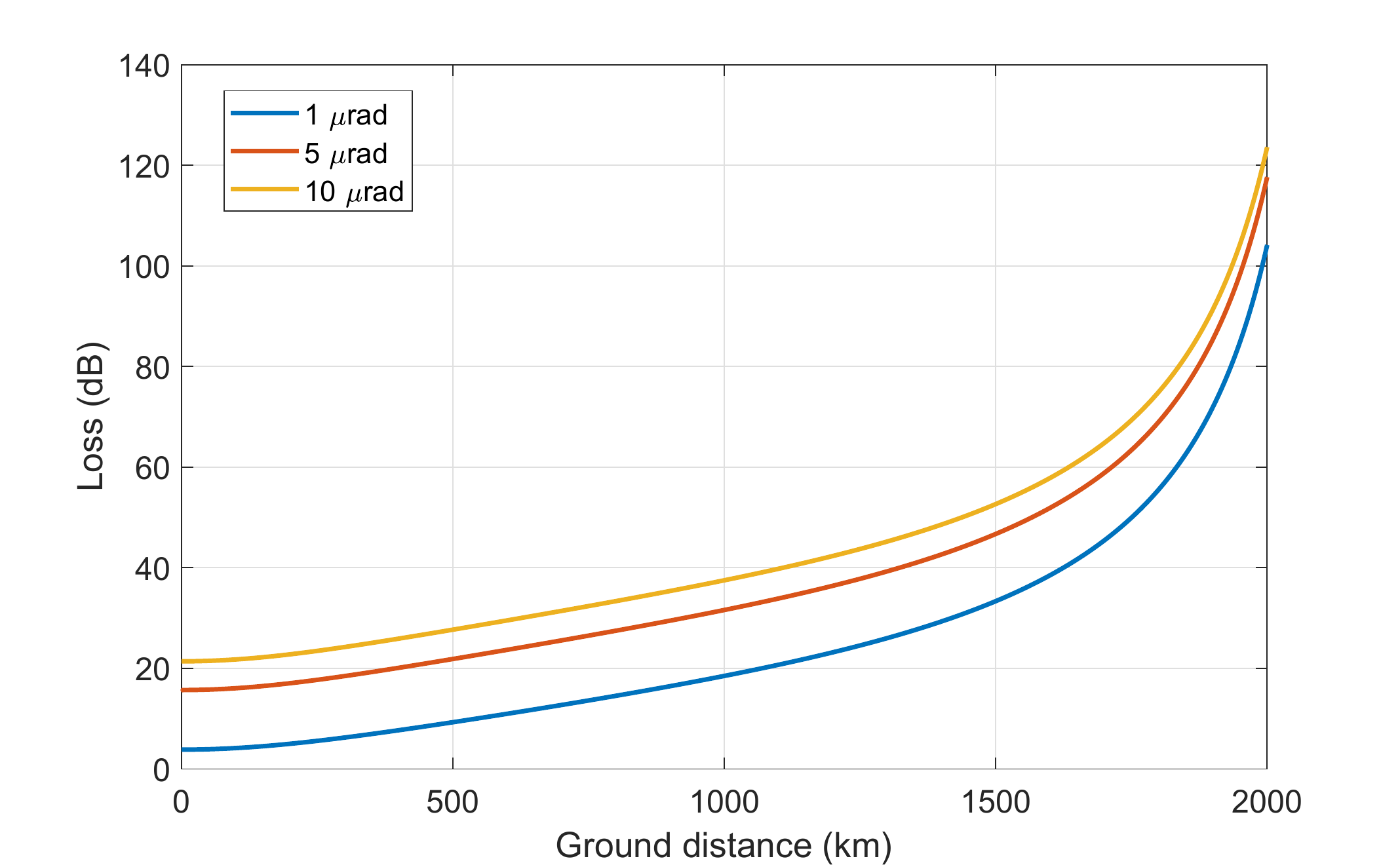} \label{fig:atmloss}}\\
\subfloat[Channel loss with satellite separation distance.]{\includegraphics[width=0.9\columnwidth]{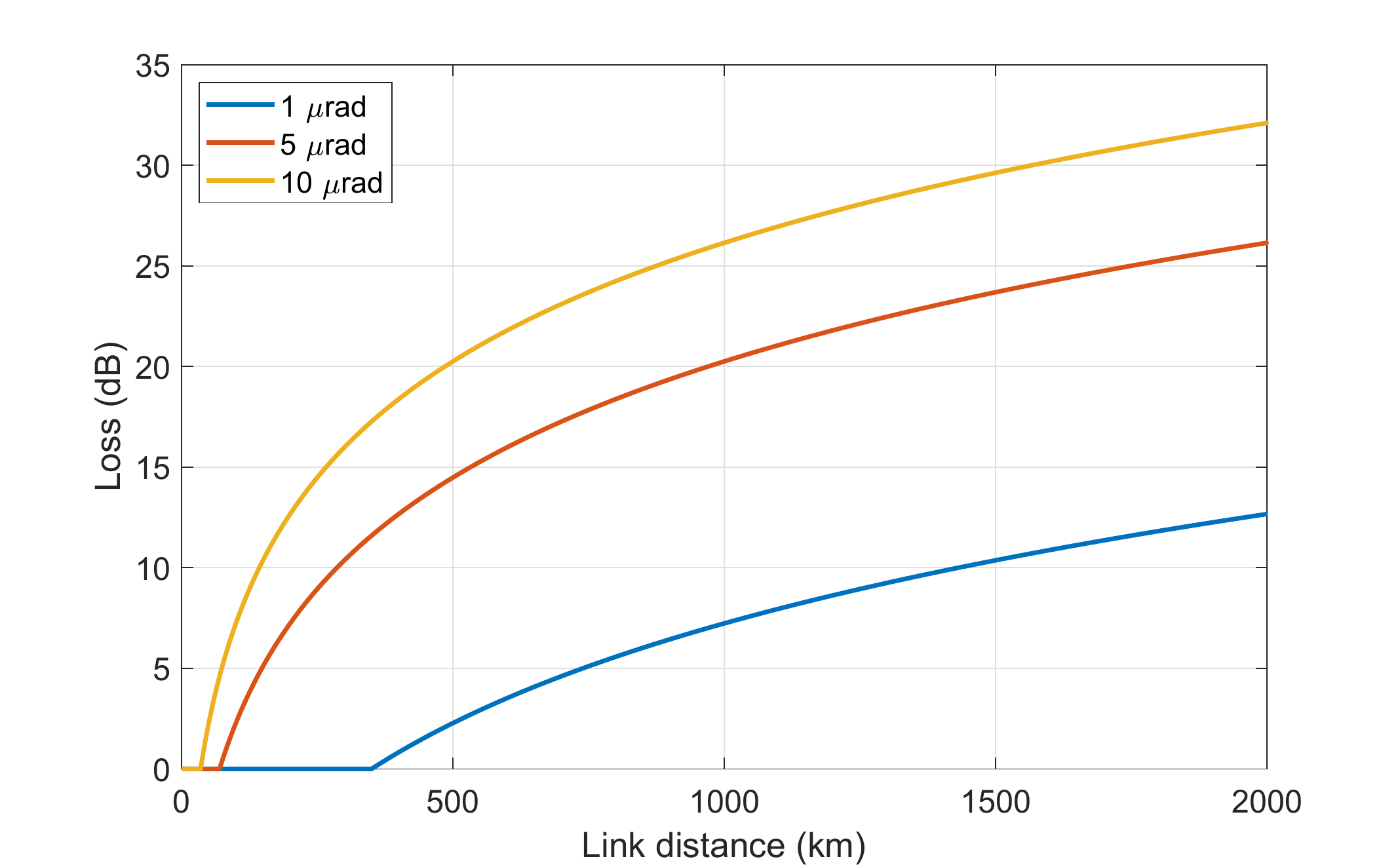}\label{fig:satloss}} \\
\caption{Optical channel losses with different beam divergences. (a) Satellite-ground connection. (b) Satellite-satellite connection. }
\label{fig:losses}
\end{figure}

{\itshape Pointing losses:} Vibration and mechanical stress due, for example to thermal dilation, cause error in the point and further loss. By assuming  that the distribution for pointing error angle follows is a Gaussian with zero mean and $\sigma_{\text{point}}$ standard deviation, this loss contribution can be modelled as \cite{2015arXiv150604836K}:
\begin{equation}
    \eta_{\text{point}}=\exp\left[-8\sigma^2_{\text{point}}/\omega_0^2\right],
\end{equation}
for a diffraction limited beam at optical wavelength a point error of $1\,\mu$rad causes decrease of the transmittance of
about 10~\%.
Fig. \ref{fig:losses} shows the channel losses when we only consider the diffractive losses and atmospheric absorption. Beam tracking errors are not included. We assume a transmitting (receiver) telescope radius of 0.15~m (0.5~m) and a low-earth orbit with $h=400$~km. Fig.~\ref{fig:atmloss} shows that atmospheric loss becomes dominant at large distances with decreasing grazing angle. Fig.~\ref{fig:satloss} shows inter-satellite losses where only diffractive losses are considered. The bigger source of noise for the current setting are given by the detectors dark counts and stray light.

{\itshape Dark counts:} For a Silicon based APD the dark count rate is estimated to be around 10 counts/s, such value could be improved by several orders of magnitude by using SNSPDs. %The detrimental effect of detector dark counts on the final key rate can be mitigated by using a narrow time acquisition window $\Delta t$, typically of about a ns.  (in our case the photons are around 1 us long, to match memory bandwidth.
In this article we assumed $p_d = 10^{-6}$ for a $\sim1~\mu$s detection window which corresponds to few Hz dark count rate.

{\itshape Stray Light:} The sources of stray light are divided in natural sources, such the moon and the stars, and the artificial one, the so called sky glow, produced buy the diffusion of light from human activities. Stray light can be decreased by spectral and time filtering. The number of stray counts in an acquisition window is given by \cite{skynoise}:
\begin{equation}
    N=\frac{\lambda }{hc}H_{\text{sky}}\Omega_{\text{FoV}}\Big(\frac{\pi D_R}{2}\Big)^2\Delta \lambda \Delta t,
\end{equation}
where $H_{\text{sky}}$ is the total sky brightness and $\Delta \lambda$  the spectral bandwidth typical of which are in the nm region. It is worth noting that, in the optical domain, the number of stray photons can vary of several orders of magnitude according to the sky condition, e.g. the presence of the Moon \cite{skynoise}.

\subsection{Keyrate calculations}

\noindent
For numerical calculations of the secret key rate, we consider a pair of quantum memories that send entangled photons to their respective end users as illustrated in Fig.~\ref{figPanayi}. The performance of this memory-assisted protocol is assessed in terms of the secret key rate achievable by the BB84 cryptographic protocol. The secret keyrate for this is lower bounded by~\cite{Panayi2014, Luong2016} 
\begin{equation}
R=\frac{Y}{2}\left[1-h\left(e_{X}\right)-f h\left(e_{Z}\right)\right],
\label{eqtn:keyrate}
\end{equation}
where $Y$ is the probability per channel use that both Alice and Bob's measurements and the Bell state measurement was successful, $e_X$ ($e_Z$) is the QBER in the $X$ ($Z$) basis, $f$ is the error correction inefficiency and $h(e)$ is the binary entropy function defined via $h(e) = -e\log_{2}e-(1-e)\log_{2}(1-e)$. Details of how $Y$ and the individual errors are calculated can be found in Refs.~\cite{Panayi2014, Trenyi2020}.
%In this article we do not consider general relativistic corrections to the photon arrival times, as calculated in~\cite{Bruschi2014}. These effects might lead to a small, but yet measurable amount of noise in the channel. 
\subsection{Temporal multimode considerations} 
\label{sec:TempMM}

\noindent
Our scheme relies on QMs to absorb many photons in a temporal multimode manner~\cite{Boone2014}. The number of required modes depends on the brightness ($R_{\mathrm{s}}$) and efficiency ($\eta_\mathrm{s}$) of the source and the maximum transmission of the channel ($\eta_{\mathrm{tr}, \max }$) and is given by
\begin{equation}N_{\mathrm{mod}}=R_{\mathrm{s}} \eta_{\mathrm{s}} \eta_{\mathrm{tr}, \max } \frac{L_{0}}{c}.\end{equation}
The protocol requires a temporal multimode storage of $N=365$ for the calculations presented in Fig.~\ref{fig:DistvsRate}.


%\bibliographystyle{ieeetr}
%\bibliography{QSAT}
\bibliography{ArxivBib}

\end{document}